%% file: ckm2012_jbrod.tex
\documentclass[12pt]{article}
\usepackage{epsfig}
\usepackage{amsmath}
\usepackage{graphicx}
\usepackage{subfigure}

\input econfmacros.tex
\def\Title#1{\begin{center} {\Large {\bf #1} } \end{center}}

\begin{document}

\Title{Direct CP violation in singly Cabibbo-suppressed D-meson
  decays}

\bigskip\bigskip


\begin{raggedright}  

{\it Joachim Brod\\
Department of Physics\\
University of Cincinnati\\
Cincinnati, OH 45221, USA}
\bigskip\bigskip
\end{raggedright}

\begin{abstract}
  The LHCb and CDF collaborations reported a surprisingly large
  difference between the direct CP asymmetries, $\Delta {\cal
    A}_{CP}$, in the $D^0 \to K^+ K^-$ and $D^0 \to \pi^+ \pi^-$ decay
  modes. We show that this measurement can be plausibly explained
  within the standard model under the assumption of large penguin
  contractions matrix elements and nominal $U$-spin breaking. A
  consistent picture arises, accommodating the large difference
  between the decay rates, and the measured decay rates of the $D \to
  K \pi$ modes.
\end{abstract}

{\it Proceedings of CKM 2012, the 7th International Workshop on the
  CKM Unitarity Triangle, University of Cincinnati, USA, 28 September
  - 2 October 2012}

\section{Introduction}

The $D^0\to K^+K^-$ and $D^0\to \pi^+ \pi^-$ decays are induced by the
weak interaction via an exchange of a virtual $W$ boson and are
suppressed by a single power of the Cabibbo angle. Direct $CP$
violation in singly Cabibbo-suppressed (SCS) $D$-meson decays is
sensitive to contributions of new physics in the up-quark sector,
since it is expected to be small in the standard model: the $b$-quark
penguin amplitudes necessary for interference are down by a loop
factor and small Cabibbo-Kobayashi-Maskawa (CKM) matrix elements, and
there is no heavy virtual top quark which could provide substantial
breaking of the Glashow-Iliopoulos-Maiani (GIM) mechanism. Naively,
one would thus expect effects of order
$\mathcal{O}([V_{cb}V_{ub}/V_{cs}V_{us}] \alpha_s/\pi) \sim
0.01\%$\footnote{Note that the possibility of $CP$ violation by tree
  amplitudes has already been pointed out in~\cite{Golden:1989qx}.}.

We define the amplitudes for final state $f$ as
\begin{equation}
\begin{split}
A_f & \equiv  A(D \to f ) = A^T_{f} \big[1+r_f
  e^{i(\delta_f-\phi_f)}\big],\\ 
\overline{A}_{f} & \equiv A(\bar D \to f ) =  A^T_{f}\big[1+r_f
  e^{i(\delta_f+\phi_f)}\big]\, .
\end{split}
\end{equation}
Here $A^T_{f}$ is the dominant tree amplitude and $r_f$ the relative
magnitude of the subleading amplitude, carrying the weak
phase $\phi_f$ and the strong phase
$\delta_f$. We can now define the direct $CP$ asymmetry as
\begin{equation}
\begin{split}
{\cal A}_f^{\rm dir} \equiv 
{|A_f|^2 -|\bar A_{f}|^2 \over |A_f|^2  + | \bar A_{ f } |^2 }  = 2
r_f \sin \gamma \sin \delta_f\, , 
\end{split}
\end{equation}
where the last equality holds up to corrections of
$\mathcal{O}(r_f^2)$. LHCb and CDF measure a time-integrated $CP$
asymmetry. The approximately universal contribution of indirect $CP$
violation cancels to good approximation in the difference
\begin{equation}
\begin{split}
\Delta {\cal A}_{CP}={\cal A}_{CP}(D\to K^+K^-) -
{\cal A}_{CP}(D\to \pi^+\pi^-)\, .
\end{split}
\end{equation}
The measurements of LHCb, $\Delta {\cal A}_{CP} = (-0.82 \pm 0.21 \pm
0.11)\%$~\cite{Aaij:2011in}, CDF, $\Delta {\cal A}_{CP}= (-0.62 \pm
0.21 \pm 0.10)\%$~\cite{Collaboration:2012qw}, and inclusion of the
indirect $CP$ asymmetry $A_\Gamma$~\cite{Aubert:2007en,Staric:2007dt},
lead to the new world average (including the
Babar~\cite{Aubert:2007if}, Belle~\cite{Staric:2008rx}, and
CDF~\cite{Aaltonen:2011se} measurements) $\Delta {\cal A}_{CP}= (-0.67
\pm 0.16)\%$~\cite{Collaboration:2012qw}. 

We show that the large difference of SCS branching ratios,
$\text{Br}(D^0\to K^+K^-) \approx 2.8 \times \text{Br}(D^0\to \pi^+
\pi^-)$, together with nominal $U$-spin breaking of
$\mathcal{O}(20\%)$, implies large penguin matrix elements, which in
turn account for the large value of $\Delta {\cal A}_{CP}$.

\section{A consistent picture}\label{sec:cp}

The starting point of our analysis is the weak effective Hamiltonian
\begin{equation}\label{eq:Heff}
\begin{split}
H_{\rm eff}^{\rm SCS} = \frac{G_F}{\sqrt{2}} & \left\{
\left(V_{cs} V_{us}^*- V_{cd} V_{ud}^*\right)  \sum_{i=1,2} C_i \left(Q_i^{\bar s s} - Q_i^{\bar dd} \right)/2 \right. \\
&\left. -V_{cb} V_{ub}^* \,\left[ \sum_{i=1,2} C_i  \left(Q_i^{\bar s s} + Q_i^{\bar d d} \right)/2  +  \sum_{i=3}^{6} 
C_i Q_i + C_{8g} Q_{8 g} \,\,\right]\right\}
+ {\rm h.c.}\, . 
\end{split}
\end{equation}
The Wilson coefficients of the tree operators $Q_1^{\bar p p'} = (\bar
p u)_{V-A} \otimes (\bar c p')_{V-A}$, $Q_2^{\bar p p'} = (\bar
p_\alpha u_\beta)_{V-A} \otimes (\bar c_\beta p'_\alpha)_{V-A}$, the
penguin operators $Q_{3\ldots 6}$, and the chromomagnetic operator
$Q_{8g}$, can be calculated in perturbation
theory~\cite{Buchalla:1995vs}. The hadronic matrix elements are of
nonperturbative nature and will ultimately can be computed using
lattice QCD~\cite{Hansen:2012tf}. We will estimate their size using
experimental data.

A leading power estimation of the ratio $r_f^\text{LP} \equiv
|A^P_f(\text{leading power})/A^T_f(\text{experiment})|$, using naive
factorization and $\mathcal{O}(\alpha_s)$ corrections, yields
$r_{K^+K^-}^\text{LP} \approx (0.01 - 0.02)\%$,
$r_{\pi^+\pi^-}^\text{LP} \approx (0.015 -
0.028)\%$~\cite{Brod:2011re}. This is consistent with, yet slightly
larger than the naive scaling estimate. We expect the signs of ${\cal
  A}_{K^+K^-}^\text{dir}$ and ${\cal A}_{\pi^+\pi^-}^\text{dir}$ to be
opposite, if $SU(3)$ breaking is not too large; so for $\phi_f =
\gamma \approx 67^\circ$ and $\mathcal{O}(1)$ strong phases we obtain
$\Delta {\cal A}_{CP} (\text{leading power}) = \mathcal{O}(0.1\%)$, an
order of magnitude smaller than the measurement. However, from $SU(3)$
fits~\cite{Cheng:2010ry,Pirtskhalava:2011va,Cheng:2012wr,Bhattacharya:2012ah,Feldmann:2012js}
we know that power corrections can be large. To be specific, we look
at insertions of the penguin operators $Q_4$, $Q_6$ into
power-suppressed annihilation amplitudes. The associated penguin
contractions of $Q_1$ cancel the scale and scheme dependence. A rough
estimate of their size
leads to $r_{f}^\text{PC}(Q_4) \approx (0.04 - 0.08)\%$,
$r_{f}^\text{PC}(Q_6) \approx (0.03 - 0.04)\%$, where
$r_{f,i}^\text{PC} (Q_i) \equiv |A^P_f(\text{power
  correction})/A^T_f(\text{experiment})|$ for the insertion of $Q_i$,
with an uncertainty of a factor of a few. Larger effects are very
unlikely~\cite{Brod:2011re}. Again assuming $\mathcal{O}(1)$ strong
phases, this leads to $\Delta {\cal A}_{CP} (r_{f,1}) =
\mathcal{O}(0.3\%)$ and $\Delta {\cal A}_{CP} (r_{f,2}) =
\mathcal{O}(0.2\%)$ for the two insertions. Thus, a standard-model
explanation seems plausible.

This conclusion receives further support from data. The large
difference of SCS branching ratios translates into a ratio of
amplitudes (normalized to phase space) of $A(D^0\to K^+K^-) \approx
1.8 \times A(D^0\to \pi^+ \pi^-)$, whereas the amplitudes would be
equal in the $SU(3)$ limit. This has often been interpreted as a sign
of large $SU(3)$ breaking. On the other hand, the ratio of the
Cabibbo-favored (CF) to the doubly Cabibbo-suppressed (DCS) amplitude
is $A(D^0\to K^-\pi^+) \approx 1.15 \times A(D^0\to K^+ \pi^-)$, after
accounting for CKM factors, in accordance with nominal $SU(3)$
breaking of $\mathcal{O}(20\%)$. This value is affirmed by the fact
that the experimental amplitudes satisfy the sum rule relation
\begin{equation}
\frac{|A(D^0 \to K^+ K^-)| + |A(D^0 \to \pi^+ \pi^-)|}{|A(D^0 \to K^+
  \pi^-)| + |A(D^0 \to K^- \pi^+)|} - 1 = (4.0\pm 1.6)\%. 
\end{equation}
This expression would vanish in the $U$-spin limit and receives
correction quadratic in $U$-spin breaking.

An inspection of the effective Hamiltonian~\eqref{eq:Heff} shows that
the combination $P$ of penguin contractions of $Q_{1,2}^{\bar s s} $
and $Q_{1,2}^{\bar dd}$ proportional to $V_{cb} V_{ub}^*$ is $U$-spin
invariant, while $P_\text{break}$, the combination of penguin
contractions contributing to the decay rates vanishes in the $U$-spin
limit. $P_\text{break}$ contributes with opposite sign to the two SCS
decay rates, and $P$ gives rise to a nonvanishing $\Delta {\cal
  A}_{CP}$. Guided by the considerations above, we perform a $U$-spin
decomposition of the amplitudes to all four (CF, SCS, DCS) decays, and
fit these amplitudes to the data (branching ratios and $CP$
asymmetries)~\cite{Brod:2012ud}. There we also provide an exact
definition of the amplitudes and a translation between the $U$-spin
decomposition and the operator picture.

\begin{figure}
\rule{5cm}{0.2mm}\hfill\rule{5cm}{0.2mm}\hfill\rule{5cm}{0.2mm}
\vskip 0.2cm
\includegraphics[scale=0.20]{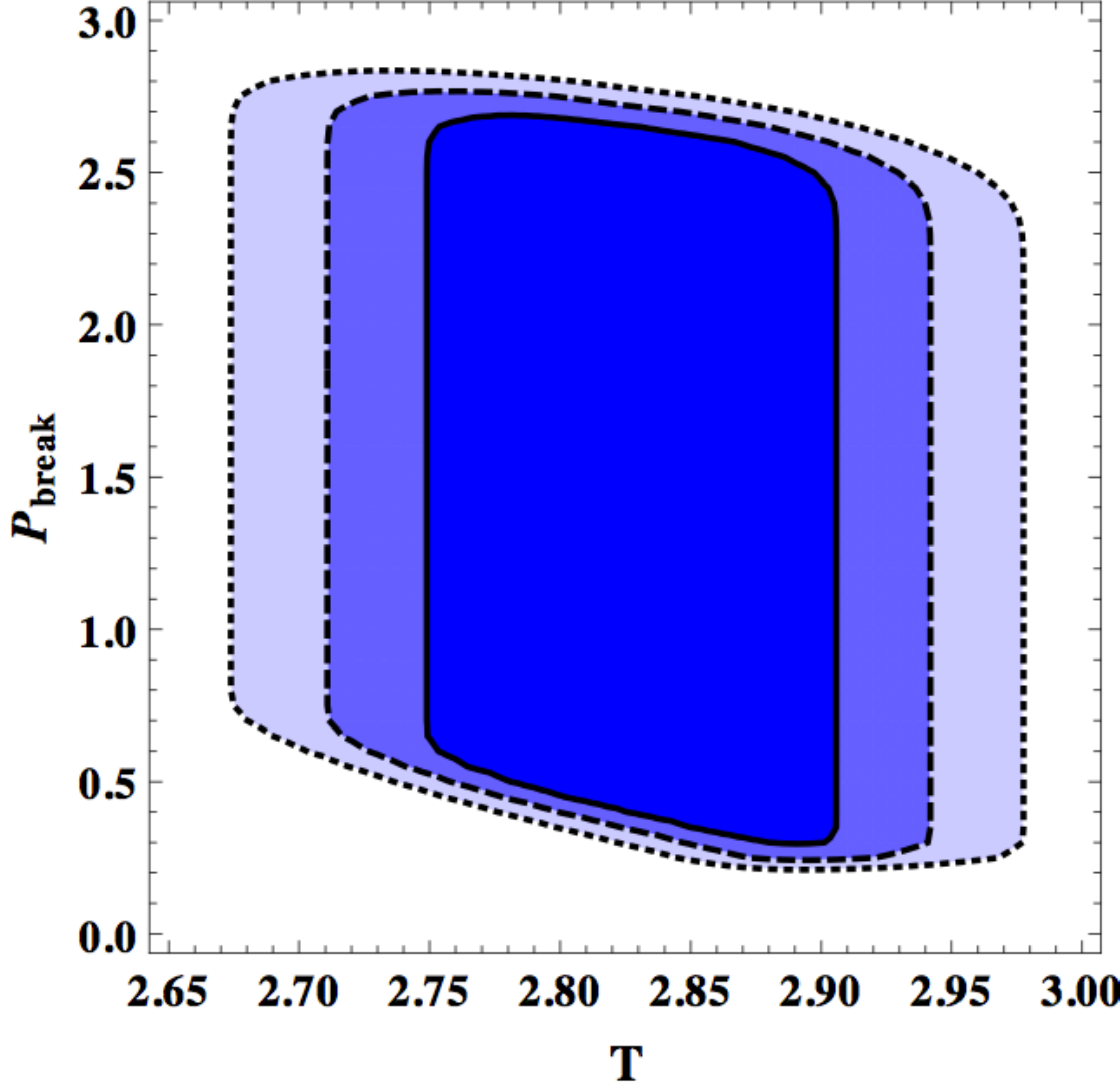}~~~~~~~~
\includegraphics[scale=0.165]{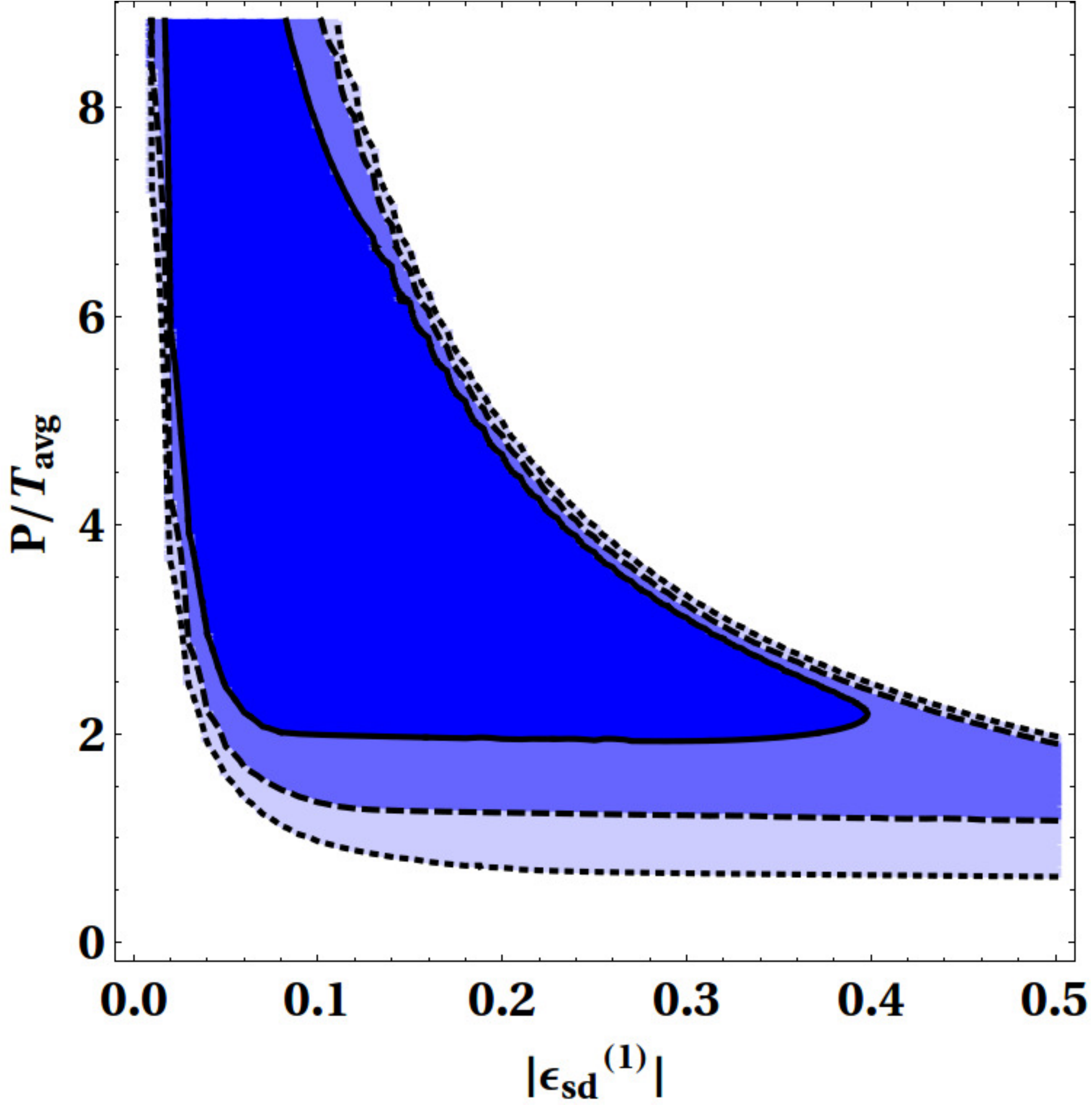}~~~~~~~~
\includegraphics[scale=0.175]{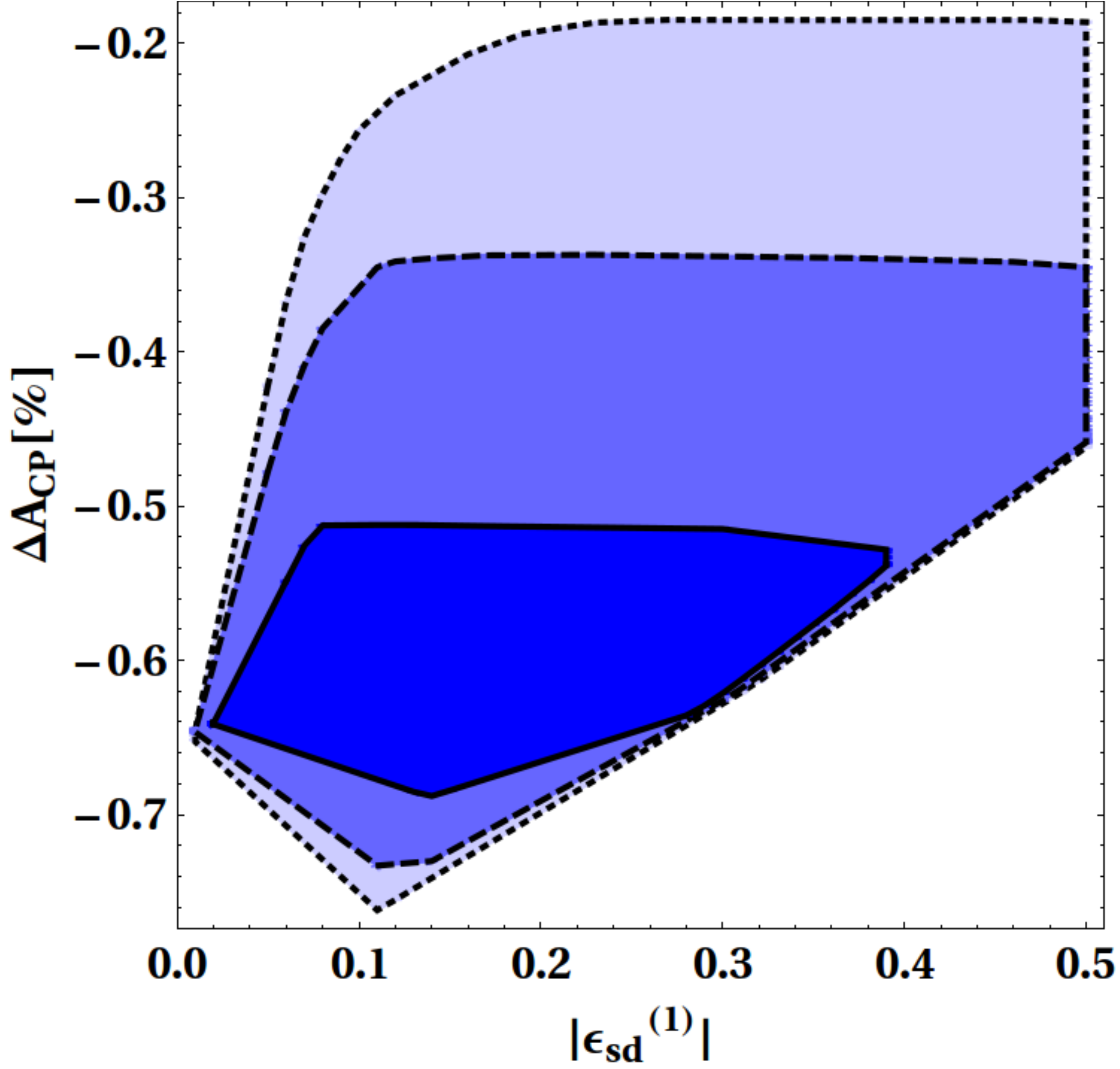}
\rule{5cm}{0.2mm}\hfill\rule{5cm}{0.2mm}\hfill\rule{5cm}{0.2mm}
\caption{The results of our fit. Solid, dashed, and dotted lines
  denote one-sigma, two-sigma, and three-sigma contours, respectively. Left
  panel: A fit to the branching ratios only yields $P_\text{break}
  \equiv \epsilon_{sd}^{(1)} P \sim T$, assuming nominal $U$-spin
  breaking. $T$ is the tree amplitude. The lower bound of
  $P/T_\text{avg}$ in the middle panel is directly related to the
  large difference of decay rates for the SCS modes. ($T_\text{avg}$
  is the average value of $T$ from the fit). It translates into the
  upper bound on $\Delta {\cal A}_{CP}$ -- the fit results can
  naturally accommodate the measured value (right panel).
  \label{fig:1}}
\end{figure}

Our main point is that under the assumption of nominal $U$-spin
breaking, a broken penguin $P_{\rm break}$, which explains the
difference of the $D^0 \to K^+K^-$ and $D^0 \to \pi^+\pi^-$ decay
rates, implies a $\Delta U=0$ penguin $P$ that naturally\footnote{An
  important side remark is that no fine tuning of strong phases is
  required~\cite{Brod:2012ud}.} yields the observed $\Delta {\mathcal
  A}_{CP}$.  The scaling $P_{\rm break } \sim \epsilon_U P$ together
with our fit to the branching ratios alone yields $P_{\rm break } \sim
T/2$ (see Fig.~\ref{fig:1}), leading to the estimate
\begin{equation} \label{randPovT} 
r_{\pi^+\pi^-, K^+K^-}\simeq
\left|\frac{V_{cb}V_{ub}}{V_{cs}V_{us}}\right|\cdot
\left|\frac{P}{T\pm P_{\rm break}}\right|\sim
\frac{|V_{cb}V_{ub}|}{|V_{cs}V_{us}|} \frac{1}{2\,\epsilon_U}\sim 0.2\%,
\end{equation}
for $\epsilon_U \sim 0.2$. This is consistent with the measured
$\Delta {\mathcal A}_{CP}$ assuming $\mathcal{O}(1)$ strong phases. A
fit to the full data set including $CP$ asymmetries confirms this
naive estimate (see Figure~\ref{fig:1}), showing that large penguin
contraction matrix elements together with nominal $U$-spin breaking
lead to a consistent picture accommodating all data on decay rates and
$CP$ asymmetries of all four (CF, SCS, DCS) modes~\cite{Brod:2012ud}.

\bigskip I thank Yuval Grossman, Alexander Kagan, and Jure Zupan for
the pleasant and fruitful collaboration, and the organizers of the
``CKM 2012 workshop'' for the invitation to this inspiring conference.
The work of J.~B. is supported by DOE grant FG02-84-ER40153.

\end{document}

%% file: econfmacros.tex
\def\beq{\begin{equation}}
\def\eeq#1{\label{#1}\end{equation}}
\def\eeqn{\end{equation}}

\def\beqa{\begin{eqnarray}}
\def\eeqa#1{\label{#1}\end{eqnarray}}
\def\eeqan{\end{eqnarray}}

\let\bar=\overbar

\def\Dslash{\not{\hbox{\kern-4pt $D$}}}
\def\dslash{\not{\hbox{\kern-2pt $\del$}}}

\def\msb{{\bar{\ssstyle M \kern -1pt S}}}

